\documentstyle[12pt,epsf,fleqn]{article}     
\newfont{\titel}{cmbx10 scaled 2200}
\newfont{\autoren}{cmti10 scaled 1728}
\newfont{\ueberschrift}{cmr10 scaled 1440}

\renewcommand{\baselinestretch}{1.2}

\begin{document}

\baselineskip30pt
{ \titel  
Fractal properties of relaxation clusters \\ 
and phase transition in a \\ 
stochastic sandpile automaton
\newline
\newline}

\baselineskip17pt

{\autoren{S.~L\"{u}beck and K.D. Usadel}\\
Theoretische Physik, Universit\"{a}t Duisburg\\
 Lotharstr.1, 47048 Duisburg, Deutschland \\
E-mail: sven@hal6000.thp.Uni-Duisburg.DE \\
\\
B.~Tadi\'c\\
Jo\v{z}ef Stefan Institute, University of Ljubljana\\ 
P.O. Box 100, 61111-Ljubljana, Slovenia \\ 
\newline\newline\newline\newline
}

\renewcommand{\baselinestretch}{1.2}

{\begin{center} \bf Abstract \end{center}}

We study  numerically the spatial properties of relaxation clusters in 
a two dimensional sandpile automaton with dynamic rules depending 
stochastically on a parameter $p$, which models the effects of static 
friction.
In the limiting cases $p=1$ and $p=0$ the model reduces to the
critical height model and critical slope model, respectively.
At $p=p_c$,  a continuous phase transition occurs to the state  
characterized by a nonzero average slope.
Our analysis reveals that the loss of finite average slope at  
the transition is accompanied by the loss of fractal properties of the 
relaxation clusters.
\\

{\begin{center} \bf Keywords \end{center}}

Self-organized criticality, sandpile automata, 
relaxation cluster, fractal dimension
\newpage

\textwidth16cm
\textheight23cm

{\ueberschrift 1\hspace{0.6cm} Introduction\\}

The term self-organized criticality (SOC) (Bak 1987, Bak 1988) 
applies to certain 
nonlinear extended dynamical systems which tend to self-organize into
a steady state with long range spatial correlations, analogous to the
critical state close to  equilibrium second-order phase transitions.
The dynamics of these systems consists of series of events (avalanches)
in which the system is repeatedly perturbed and let to relax according 
to certain microscopic dynamic rules. After a certain large number of time
steps, the system "learns" its response to the external perturbation.
Clusters of relaxed sites have been shown to have fractal properties
(J\'anosi 1994).
Therefore, the study of spatial properties of  relaxation clusters
offers an additional  possibility to analyze the character of the
dynamical process itself.

\hspace{0.5cm}Sandpile automata models (Bak 1987, Dhar 1989, Manna 1991) 
are well known prototype 
models exhibiting SOC. By adding a particle from the outside, a sandpile is
perturbed and  the perturbation may lead to instabilities at neighboring
sites if the local height of sand exceeds a critical value $h_c$ --- critical 
height model (CHM) (Bak 1987, Dhar 1989), or if the local slopes exceed a 
critical value $\sigma _c$ --- critical slope model (CSM) (L\"ubeck 1993, Manna 1991).
The self-tuning of the system in the case of the CHM is accompanied with
a state of a zero average slope. A finite average slope characterizes
a CSM.  

\hspace{0.5cm}In the present work we study a two dimensional sandpile-type automaton 
with preferred direction  
in which the updating rules are tuned continuously by changing 
a parameter $p$  between CHM (for $p=1$) and CSM (for $p=0$). 
For all values $0 < p< 1$ the dynamics is stochastic.
According to these dynamic rules, the relaxation clusters may have holes and
dendritic forms. 
We  present the results of numerical investigations 
of the spatial properties, i.e. distributions of  size and length, and fractal
dimension of the relaxation clusters. 
Our 
conclusion is that there is a treshold value of the probability parameter, 
$p=p_c$, above which the system behaves like a CHM, characterized by
zero net slope. 
However, for  $p < p_c$ the dynamics leads to the 
state with a nonzero net slope.
The phase transition between these two states
of the automaton is characterized by a continuous appearance of the average
slope at $p \leq p_c$. 
At the same time, on approaching $p=p_c$ from above, the fractality 
of the relaxation clusters disappears.\vspace{0.5cm} \\

\newpage
{\ueberschrift 2\hspace{0.6cm}Model and Numerical Simulations\\} 

We consider a two-dimensional sandpile model on a square lattice
of size $L \times L$ and integer variables $h(i,j)$, representing the local
height.
We assume a directed dynamics, i.e. particles are restricted
to flow in the downward direction (increasing $i$).
According to the widespread 'sandpile language' the first row $(i=1)$
and the last row $(i=L)$ represent the top and the bottom of the
pile, respectively.
To minimize the influence of the horizontal boundaries we limit our investigations
to periodic boundary conditions in this horizontal direction
($j$-direction).
Any site of the lattice has two downward and two upward next neighbours,
namely 
\begin{equation}
h(i+1,j_{\pm}) \hspace{0.5cm} \mbox{and} \hspace{0.5cm} h(i-1,j_{ \pm})
\hspace{0.5cm}  \mbox{with}  \hspace{0.5cm}
j_{\pm}=j \pm \frac{(1 \pm (-1)^i)}{2}.
\end{equation}
We perturb the system by adding particles at a random place on the top
of the pile according to
\begin{equation}
h(1,j) \, \mapsto \, h(1,j)+1\, , \hspace{1cm} \mbox{with random }j.
\end{equation}

A site is called unstable if the height $h(i,j)$ or at least one of the 
two slopes 
\begin{equation}
\sigma (i,j_{\pm})=h(i,j)-h(i+1,j_{\pm})
\end{equation}
 exceeds a critical value, i.e
if $h(i,j) \ge h_c$ or $\sigma (i,j_{\pm}) \ge \sigma_c$, respectively.
  
\hspace{0.5cm}In the case of the critical slope condition toppling takes place until 
both slopes become subcritical. 
In this toppling process  particles drop alternative to the downward next neighbours 
if both slopes are unstable. 
If only one slope exceeds $\sigma_c$ particles drop to the corresponding
downward neighbour.

\hspace{0.5cm}In contrast to the critical slope the critical height conditions has
a stochastic character.
If the local height exceeds the critical value $h_c$ 
toppling occurs only with the probability $p$, and then two particles drop
to the two downward neighbours.

\hspace{0.5cm}Because each relaxing cell changes the heights and slopes of its four next
neighbouring sites the stability conditions are applied at these four 'activated' 
sites in the next updating step.
Toppling may take place after adding a particle on the first row. 
We first apply the slope and then the height stability condition in parallel
for each activated site.
An avalanche stops if all activated sites are stable.
Then we start again according to Eq. (2).

One can interpret $1-p$ as being 
due to  static friction between the sand grains, which prevents toppling 
even if the height exceeds the critical value.
In the case  $p=1$,  our model is identical to the model of Dhar and 
Ramaswamy (Dhar 1989), which exhibits a robust SOC behavior.
In the  limit $p=0$,   our directed rules lead to a steady state where all 
 slopes are equal, i.e., $\sigma(i,j_{\pm})=\sigma_{c}-1$. 
Thus, if a particle is added at the first row it  
performs a directed
random walk downward the pile until it reaches the boundary and
drops out of the system --- no SOC  can  occur in this limit.

\hspace{0.5cm}We fix the critical height $h_{c}$
and the critical slope $\sigma_{c}$ 
and restrict ourselves to $h_{c}=2$ and $\sigma_{c}=8$,
although it should be emphasized that the results depend on the choice of 
these parameters.\\

\begin{figure}[h]
 \begin{center}
 \begin{minipage}{11.0cm}
 \epsfxsize=14.0cm
 \epsfysize=14.0cm
 \epsffile{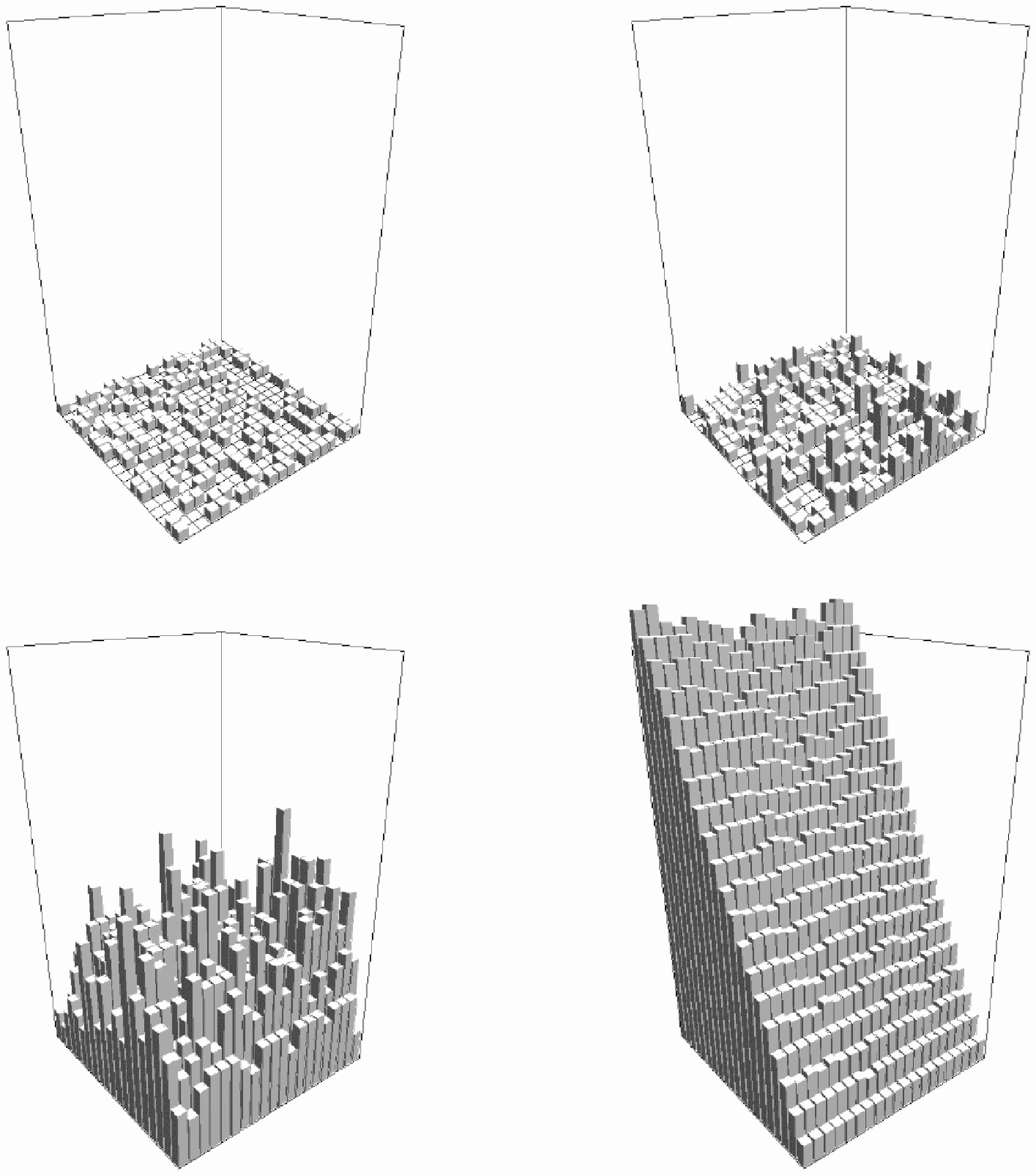} 
 \end{minipage}
 \end{center}
\end{figure}

\vspace{-1cm} 
{\bf Figure~1}\hspace{0.5cm}{Snapshots of sandpiles for $p=1$ (upper left), 0.7 (upper
              right), 0.3 (lower left), and 0.1. In all cases the upper left boundary
	      corresponds to the first row of the pile ($i=1$).}\\

\hspace{0.5cm}Starting from an empty lattice 
of linear size $L$ we add approximately $L^3$ particles to equilibrate the
system before taking measurements.
In each case the stationarity of the system  was checked by determining
that the average height $\mathopen<h(i,j)\mathclose>$ 
has reached a constant value.\\ \\

{\ueberschrift 3\hspace{0.6cm}Results\\} 

In Figure~1 we show four snapshots of the sandpile for different values of $p$. 
The upper left picture represents the pile in a steady state of the pure 
CHM ($p=1$). Only the heights 0 and 1 are present in the critical state.
With decreasing $p < 1$ a certain number of heights with values 
$h(i,j) > h_{c}$  remain in the interior of an avalanche leading to a rough
surface of the pile (see e.g. the case $p=0.7$ of Figure~1).

\hspace{0.5cm}For still lower values of $p$ the roughness
becomes more dramatic and eventually for $p=0.3$
(lower left) the surface
shows large fluctuations, indicating a change of the system behavior.
In the case  $p=0.1$, displayed in  the lower right picture, 
the heights grow up to the full size required by the CSM rules with 
fluctuations around the well defined average slope.
We now analyze this behavior quantitatively by considering
the average height $\langle h(i)\rangle$ as a function of distance from 
the top row $i$, defined as
\begin{equation}
\langle h(i)\rangle \, = \, L^{-1} \sum^L_{j=1} \, \langle h(i,j)\rangle,
\label{avh}
\end{equation}
where $\langle h(i,j)\rangle$ is determined as the average over total 
number of time steps.
In the interior of the pile 
the average height is independent of $i$ for all values of $p \geq 0.3$, 
i.e. $\langle h(i) \rangle = \langle h \rangle$, 
where $\langle h \rangle$ depends on the probability parameter $p$. 
The normalized average heights are plotted in Figure~2a vs. the distance $i$. 
Except of the close vicinity of the boundaries
the curves corresponding to different values of $p>0.3$ collapse into a single 
curve $\langle h(i) \rangle /  \langle h \rangle=1$.

\hspace{0.5cm}However, this is no longer
the case for $p < 0.3$, where the deviations from $ \langle h \rangle$ 
characteristic for the behavior at the boundaries proliferates into the interior
of the pile. 
Instead of constant values of the average heights we find for $p < 0.3$
constant values of the average slopes, which are defined as 
$\sigma(i) = \langle h(i) \rangle - \langle h(i+1) \rangle $.
In Figure~2b the normalized average slopes are shown. Except of the deviations
due to the boundaries all curves more or less collapse to 
$\sigma(i) / \sigma(i=50)  = 1$.
Simulations for some values of $p$ and different automata sizes 
($L=50,100,200$) show that these deviations are independent of $L$, i.e.
they are true boundary effects.
In the interior part of the automaton the average slopes are
independent of $i$.

\hspace{0.5cm}For  $p < 0.3$, the slopes obey a translation invariance similar to
that of the heights in the opposite limit.
The average slopes become a function of $i$ if the probability
parameter $p$ is close to $p=0.3$. 

\newpage

\begin{figure}[h]
 \begin{center}
 \epsfxsize=10.0cm
 \epsfysize=7.5cm
 \begin{minipage}{11.0cm}
 \epsffile{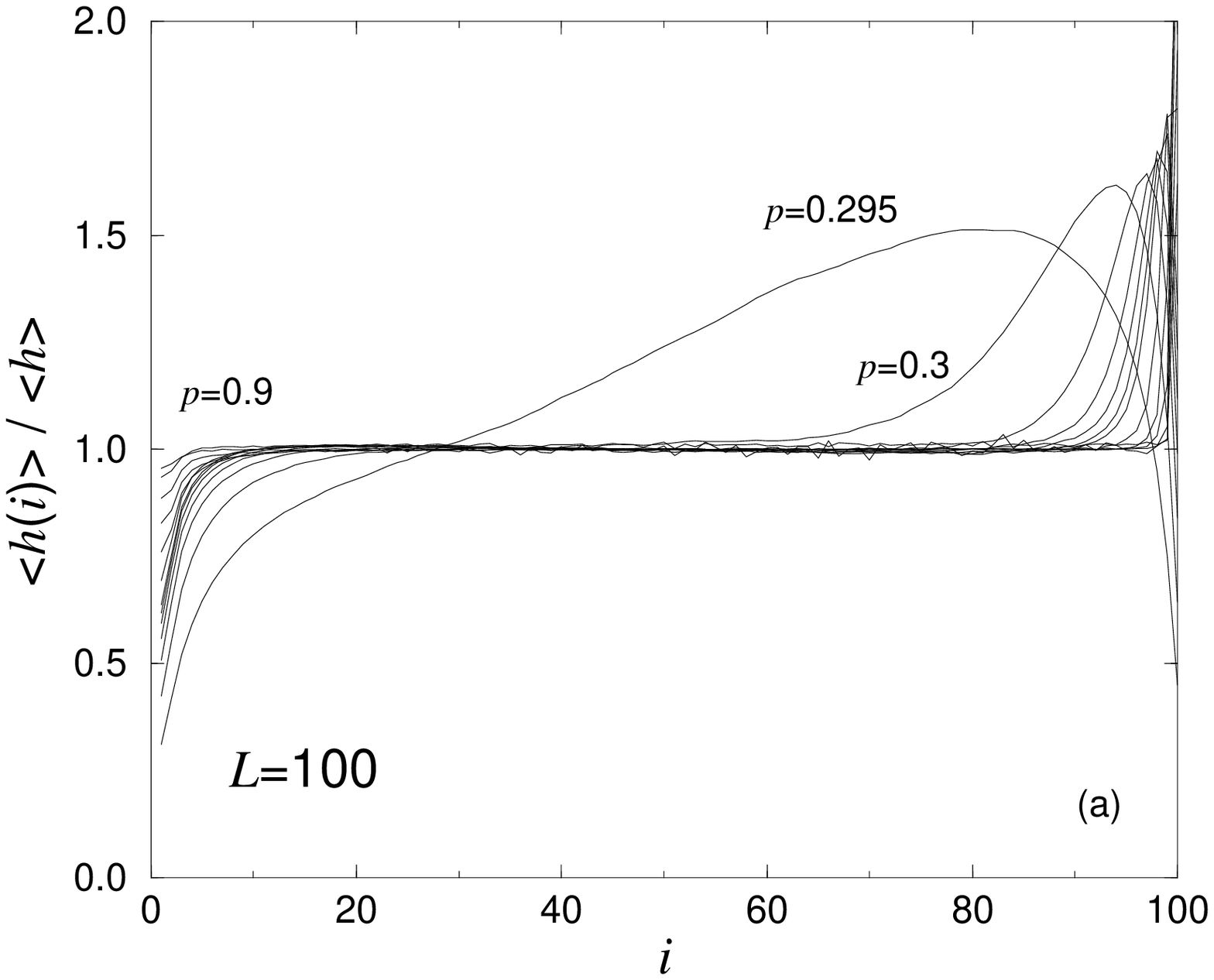}
 \vspace{-1.0cm}
 \epsfxsize=10.0cm
 \epsfysize=7.5cm
 \epsffile{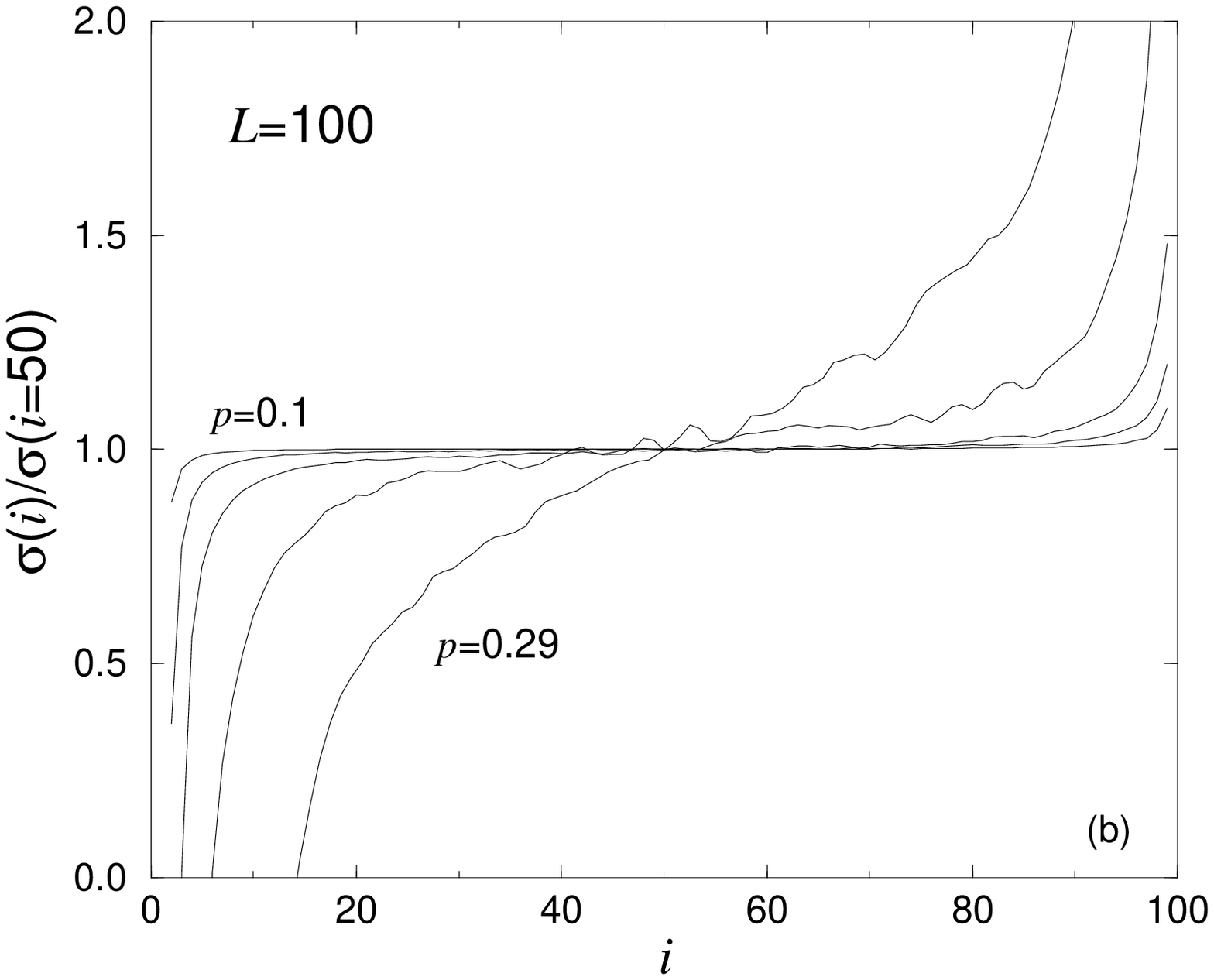}
 \end{minipage}
 \end{center}
\end{figure}

\vspace{-1cm} 
{\bf Figure~2}\hspace{0.5cm}{(a) Normalized average height $\langle h(i) \rangle$ at 
 distance $i$ from the top of the pile plotted vs. the distance $i$ for 
various values of $p$ between $0.9$ (flat curve) and 0.295 (curve with 
profile). (b)~Normalized average slope $\langle \sigma (i) \rangle$ for
various values of $p$ between $0.1$ (flat curve) and 0.29 (curve with profile)}.\\

\hspace{0.5cm}Since a zero net slope characterizes a CHM and a non zero net slope
a CSM, respectively, we interpret the average slopes 
\begin{equation}
\langle \sigma \rangle \, = \frac{1}{L-40} \, 
\sum_{i=20}^{L-20} \,  \, \langle \sigma(i) \rangle 
\label{netsigma}
\end{equation}
as an order parameter, which describes the transition from the
critical height to the critical slope regime.
In this definition we cut off the first and last 20 values in order to minimize
the influence of the boundaries.
The results for $\langle \sigma \rangle$ which are obtained in this way are
independent of $L$ as additional simulations for $L=200$ have shown.
Of course it is also irrelevant if the cut off length is larger than 20.

\hspace{0.5cm}The true nature of the transition as 
well as the transition point $p_c$ should be determined from the 
$p$-dependence of the average slope itself in the vicinity of $p_c$.
We determined the average net slope according to Eq.~(\ref{netsigma})
for different values of $p$. 

\hspace{0.5cm}The results are shown in Figure~3.
As the average net slope plays the role of an order parameter
in our model, we see from Figure~3 that a  
continuous transition occurs between the two regimes of the system. 
Assuming the following form 
\begin{equation}
\langle \sigma \rangle \, \sim \, (p_c-p)^\alpha , 
\label{op}
\end{equation}
we determine the transition point $p_c=0.293 \pm 
0.002$ and the exponent $\alpha = 0.8 \pm 0.05$ from the data in Figure~3.

\begin{figure}[h]
 \begin{center}
 \begin{minipage}{11.0cm}
 \epsfxsize=11.0cm
 \epsfysize=8.0cm
 \epsffile{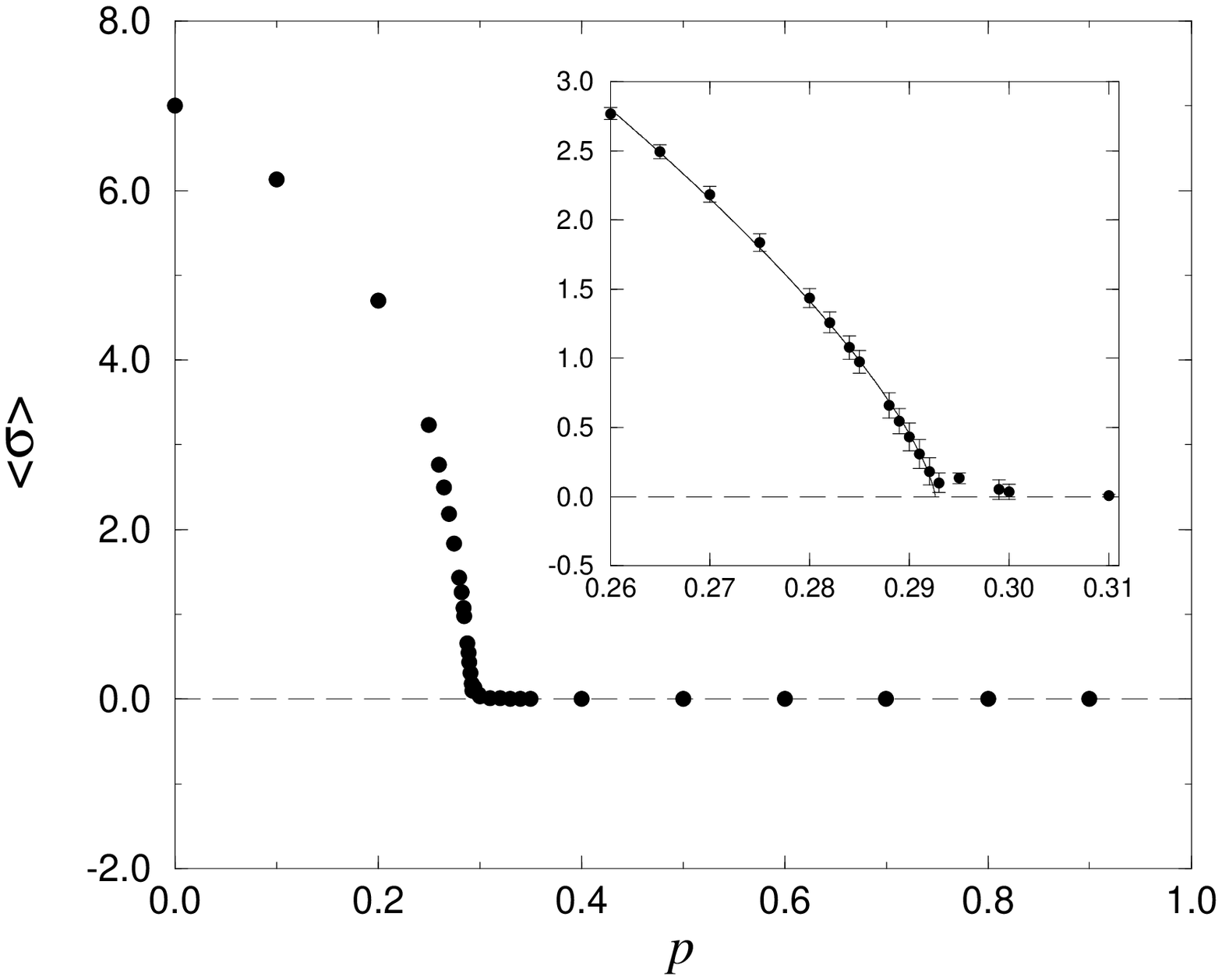}
 \end{minipage}
 \end{center}
\end{figure}

\vspace{-1cm} 
{\bf Figure~3}\hspace{0.5cm}{Average slope of the pile $\langle \sigma \rangle$
       plotted vs. probability $p$. The solid line of the inset corresponds
       to a fit according to equation (\ref{op}) with $p_c=0.293$ and 
       $\alpha=0.8$.}\\

\hspace{0.5cm}Next we consider the spatial structure of the avalanches.
One relaxation cluster consists of all sites that topple during one event
following one particle being added at the top row.
In Figure~4 we show  six representative snapshots of the relaxation clusters
taken  from simulations for different values of $p$. 
For $p=1$ the avalanches are compact.
With decreasing $p$ some supercritical sites may remain in the interior of 
the avalanche due to the stochastic character of the dynamic rules,
leading to holes and branching of the clusters. 
For  values of $p$ which are lower than  $p_c \approx 0.3$ 
this structure of the avalanches is lost. In this region the 
net nonzero slope is developed and the probability of triggering an
after-avalanche increases dramatically. The system relaxes  occasionally
through huge avalanches involving almost all sites.
In the limiting case of $p=0$ the system shows rather simple behavior,
since the rules of the CSM are deterministic and fluctuations arrond
the average slope $\sigma _c-1$ are absent.
Each added particle performs a random walk down the whole automaton.

\vspace{-1cm} 

\begin{figure}[h]
 \begin{center}
 \begin{minipage}{11.0cm}
 \epsfxsize=13.0cm
 \epsfysize=16.9cm
 \epsffile{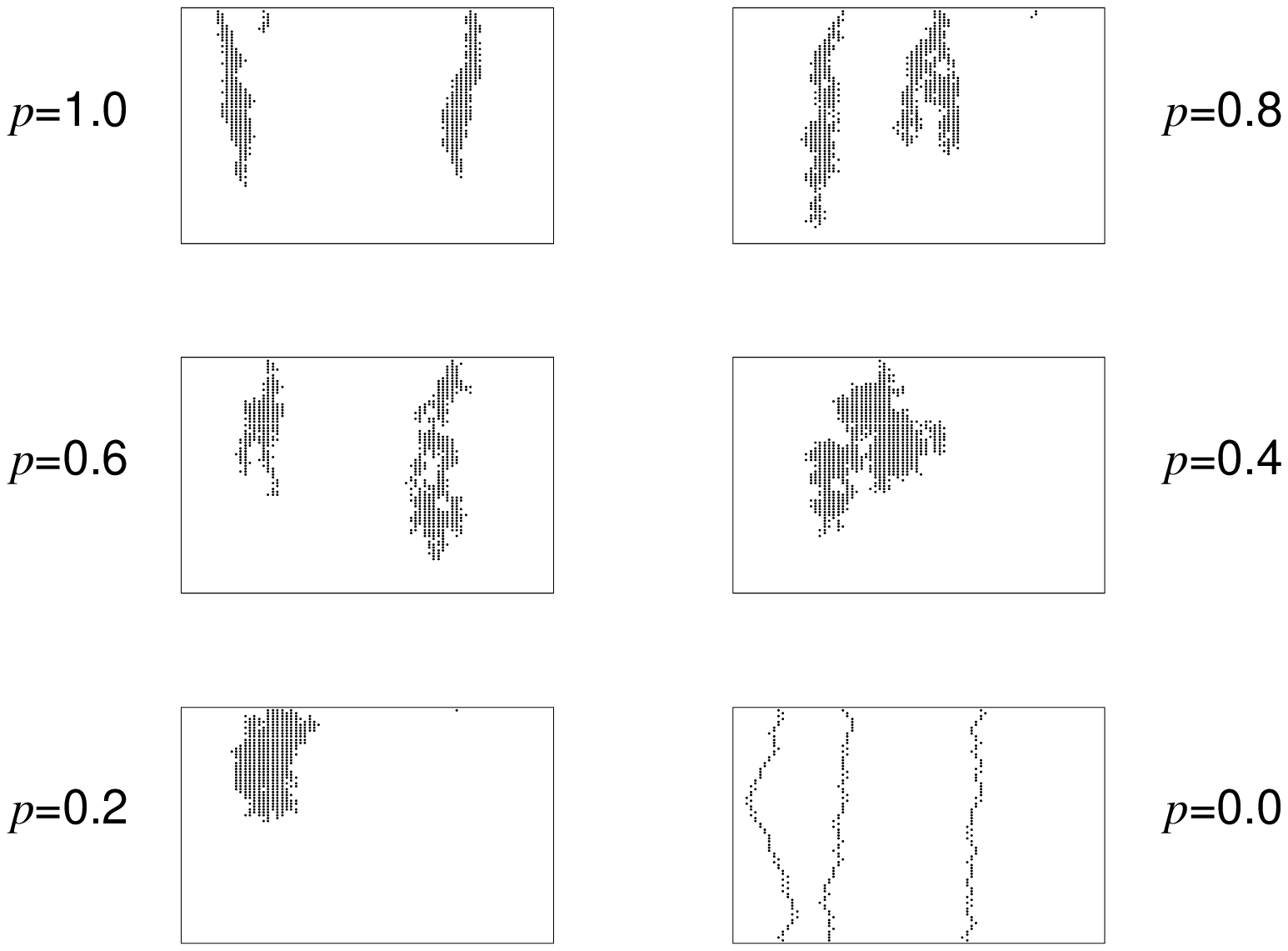} 
 \end{minipage}
 \end{center}
\end{figure}
     
\vspace{-2cm} 
{\bf Figure~4}\hspace{0.5cm}{Snapshots of relaxation clusters for a few 
values of $p$ for a linear system size of $L=80$.
Sites which toppled during an avalanche are marked as black. The base of
every picture represents the bottom of the pile ($i=L$).}\\

\hspace{0.5cm}We next analyze the scaling properties of the clusters, which we expect
to reflect these observations.
We calculate the average size of all clusters of length $l$, 
$\langle s\rangle (l)$, which scales as a power of the length $l$, i.e.,
\begin{equation}
\langle s \rangle (l) \, \sim \, l^{d_{\parallel}} \; . \label{dfr}
\end{equation}
\hspace{0.5cm}We emphasize that this definition of $d_{\parallel}$ describes only the 
scaling along the prefered direction. Actually the avalanches 
in our model are self-affine fractals, i.e. the fractal dimension along the
perpendicular direction $d_{\perp}$ is not equal to $d_{\parallel}$.
A  detailed discussion of this anisotropic behavior will be published
elsewhere.

\begin{figure}[h]
 \begin{center}
 \begin{minipage}{11.0cm}
 \epsfxsize=10.0cm
 \epsfysize=7.5cm
 \epsffile{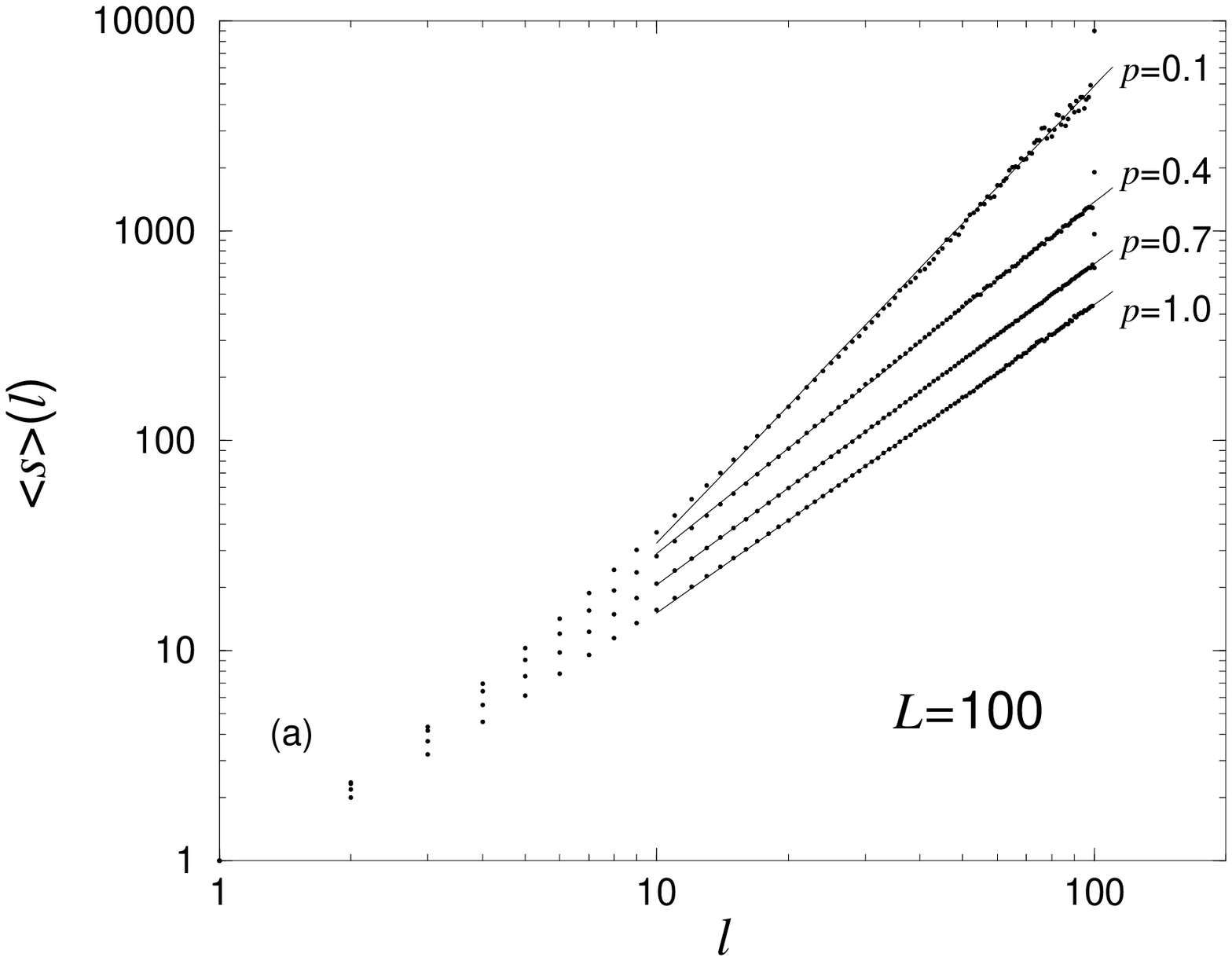}
 \vspace{-1.0cm}
 \epsfxsize=10.0cm
 \epsfysize=7.5cm
 \epsffile{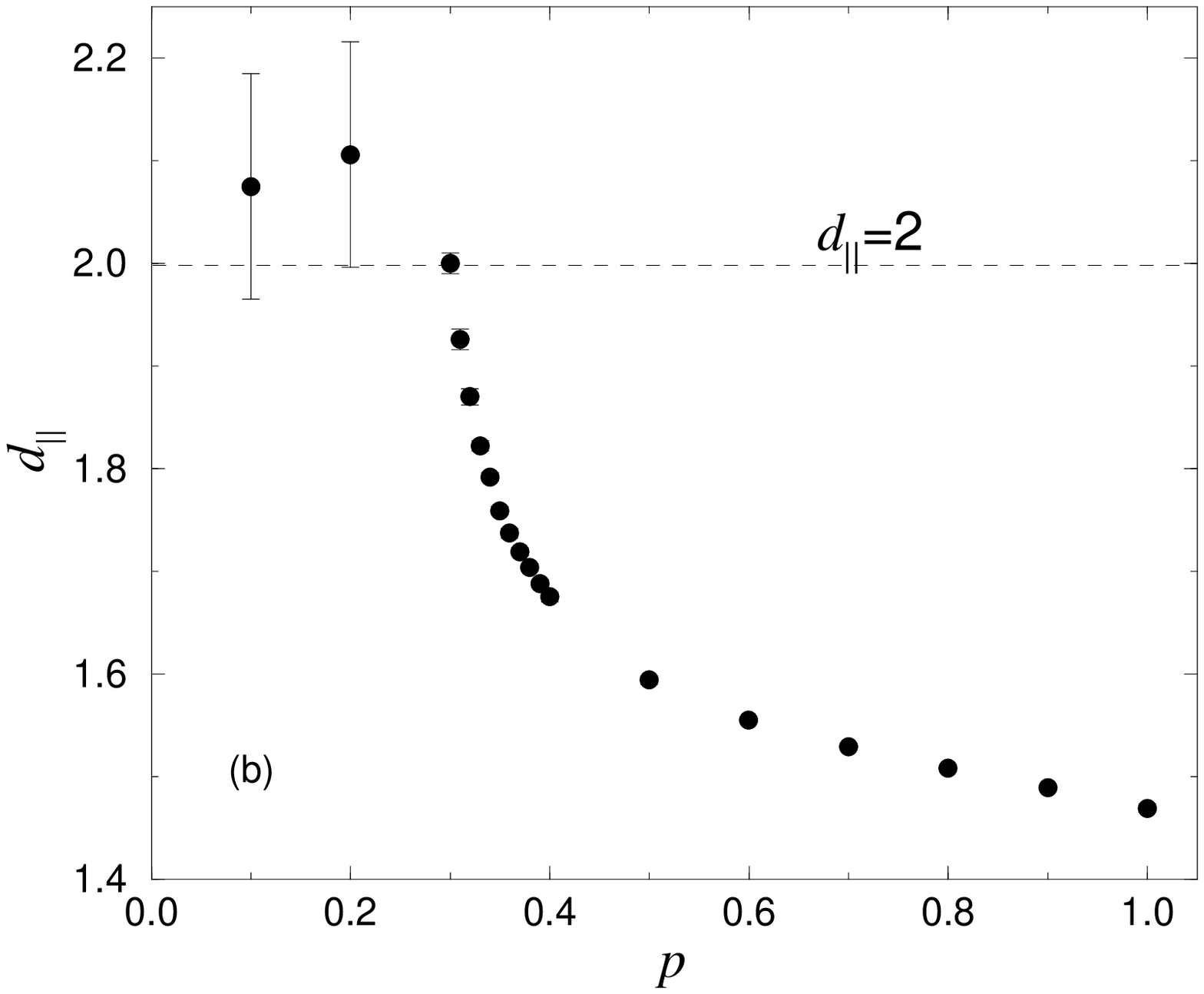}
 \end{minipage}
  \end{center}
\end{figure}

\vspace{-1cm} 
{\bf Figure~5}\hspace{0.5cm}{(a) Double logarithmic plot of the average size of 
clusters $\langle s \rangle (l)$ vs. length $l$ for various values
of $p$ as indicated.
(b)~Parallel fractal dimension $d_{\parallel}$ of the relaxation clusters
plotted vs. probability $p$. For $p \geq 0.33$ the error-bars are smaller
than the symbols.}\\

\hspace{0.5cm}The exponent in Eq.~(\ref{dfr}), which determines the scaling properties
of relaxation clusters, depends on the parameter $p$ (see  Figure~5).
We examined this $p$-dependence of the fractal dimension intensively, 
the obtained results are shown in Figure~5.
In agreement with the pictures of the avalanche shapes in Figure~4 $d_{\parallel}$ 
increases with decreasing $p$.
From $d_{\parallel}=1.469$ for $p=1$, which is in a good agreement with
the known  value $d_{\parallel}=\frac{3}{2}$ for the pure directed CHM 
(Tadi\'c 1992), the fractal  dimension grows with decreasing $p$ 
and reaches its maximum value $d_{\parallel}=2$ for $p\leq 0.3$. 
In this way the critical height regime is accompanied by a self-affine 
fractal structure of the relaxation cluster.
This fractal behavior vanishes below the transition point $p_c$ 
where the critical slope condition rules the dynamics.\\ \\

{\ueberschrift 4\hspace{0.6cm}Conclusions\\} 

In conclusion, we have shown that our stochastic sandpile automaton
exhibits a continuous phase transition in its steady state, as the 
probability parameter $p$ is varied through the transition point $p_c =
0.293$. 
The steady states for $p < p_c$ are 
characterized by a finite net average slope. 
For $p > p_c$ the average slope remains zero.
Fractality of the relaxation clusters, measured by the deviation of 
their fractal dimension $d_{\parallel}$ from the Euclidean dimension $d=2$, 
is shown to vanish at the transition point, strongly suggesting that the 
fractal character of the relaxation clusters is closely related to the 
CHM regime. 

\hspace{0.5cm}The analysis of the avalanche distributions and especially the question
whether the model displays SOC behavior for $p<1$ remains beyond the scope
of the present paper and will be published elsewhere.\\ \\

{\ueberschrift Acknowledgments\\}

This work was supported in part by the Deutsche Forschungsgemeinschaft
through Sonderforschungsbereich 166, Germany,  and by the Ministry of Science 
and Technology of the Republic of Slovenia. 
We would like to thank U.~Nowak for critical reading the manuscript.
One of us (B.T.) wishes to acknowledge the hospitality at the University of 
Duisburg, where this work was initiated. \\ \\

\newpage

{\ueberschrift References\\}

Bak~P., Tang~C. and Wiesenfeld~K. (1987) Self-Organized Criticality: 
An Explanation of  

\hspace{0.5cm}$1/f$ Noise, {\it Physical~Review~Letters}, {\bf 59}, 381.

Bak~P., Tang~C. and Wiesenfeld~K. (1988) Self-organized criticality,

\hspace{0.5cm}{\it Physical~Review~A}, {\bf 38}, 364.

Dhar~D. and Ramaswamy~R. (1989) Exactly Solved Model of
Self-Organized Critical 

\hspace{0.5cm}Phenomena, {\it Physical~Review~Letters}, {\bf 63}, 1659.

J\'anosi~I.M. and Csir\'ok~A. (1994) Fractal clusters and self-organized
criticality,

\hspace{0.5cm} {\it Fractals}, {\bf 2}, 153.

L\"ubeck~S., Usadel~K.D. (1993) SOC in a class of sandpile models
with stochastic 

\hspace{0.5cm}dynamics, {\it Fractals} {\bf 1}, 1030.

Manna~S.~S. (1991) Critical exponents of the sand pile models in
two dimensions,

\hspace{0.5cm}{\it Physica~A} {\bf 179}, 249.

Tadi\'c~B., Nowak~U., Usadel~K.D., Ramaswami~R., and 
Padlewski~S. (1992) 

\hspace{0.5cm} Scaling behavior in disordered sandpile automata, 
{\it Physical~Review~A}, {\bf 45}, 8536.

\end{document}